\documentclass{PoS}
\usepackage{axodraw}
\def\sign(#1){(\!-\!1)^{#1}}
\def\plus{\!+\!}
\def\minus{\!-\!}
\def\mydot{\!\cdot\!}
\def\nn{\nonumber \\ &&}

\def\TAB(#1,#2,#3,#4,#5,#6){
        \raisebox{-19.1pt}{
        \SetScale{0.5} \SetPFont{Helvetica}{14}
        \hspace{-15pt}
        \begin{picture}(50,39)(0,-4)
        \SetColor{Blue}
        \CArc(40,35)(25,90,270) \CArc(60,35)(25,270,90)
        \Line(40,60)(60,60) \Line(40,10)(60,10) \Line(50,10)(50,60)
        \Line(0,35)(15,35) \Line(85,35)(100,35)
        \SetColor{Black}
        \PText(53,39)(0)[lb]{#5} \PText(53,36)(0)[lt]{#6}
        \PText(35,62)(0)[rb]{#1} \PText(65,62)(0)[lb]{#2}
        \PText(65,12)(0)[lt]{#3} \PText(35,12)(0)[rt]{#4}
        \SetColor{Red}
        \SetWidth{3}
                \Line(50,10)(50,60)
                \Vertex(50,60){1.3}
                \Line(40,60)(50,60)
                \CArc(40,35)(25,90,180)
        \SetWidth{0.5}
        \end{picture}
        \SetScale{1.0}
        \hspace{-7pt}
        }
}
\def\TAD(#1,#2,#3,#4,#5,#6){
        \raisebox{-19.1pt}{
        \SetScale{0.5} \SetPFont{Helvetica}{14}
        \hspace{-15pt}
        \begin{picture}(50,39)(0,-4)
        \SetColor{Blue}
        \CArc(40,35)(25,90,270) \CArc(60,35)(25,270,90)
        \Line(40,60)(60,60) \Line(40,10)(60,10) \Line(50,10)(50,60)
        \Line(0,35)(15,35) \Line(85,35)(100,35)
        \SetColor{Black}
        \PText(53,39)(0)[lb]{#5} \PText(53,36)(0)[lt]{#6}
        \PText(35,62)(0)[rb]{#1} \PText(65,62)(0)[lb]{#2}
        \PText(65,12)(0)[lt]{#3} \PText(35,12)(0)[rt]{#4}
        \SetColor{Red}
        \SetWidth{3}
                \Line(50,10)(50,60)
        \SetWidth{0.5}
        \end{picture}
        \SetScale{1.0}
        \hspace{-7pt}
        }
}
\def\TAF(#1,#2,#3,#4,#5,#6,#7){
        \raisebox{-19.1pt}{
        \SetScale{0.5} \SetPFont{Helvetica}{14}
        \hspace{-15pt}
        \begin{picture}(50,39)(0,-4)
        \SetColor{Blue}
        \CArc(40,35)(25,90,270) \CArc(60,35)(25,270,90)
        \Line(40,60)(60,60) \Line(40,10)(60,10) \Line(50,10)(50,60)
        \Line(0,35)(15,35) \Line(85,35)(100,35)
        \SetColor{Black}
        \PText(53,39)(0)[lb]{#5} \PText(53,36)(0)[lt]{#6}
        \PText(40,65)(0)[rb]{#1} \PText(15,48)(0)[rb]{#7} \PText(65,62)(0)[lb]{#2}
        \PText(65,12)(0)[lt]{#3} \PText(35,12)(0)[rt]{#4}
        \SetColor{Red}
        \SetWidth{3}
                \Line(50,10)(50,60)
                \Line(40,60)(50,60)
                \CArc(40,35)(25,90,130)
                \Vertex(50,60){1.3}
        \SetWidth{0.5}
        \end{picture}
        \SetScale{1.0}
        \hspace{-7pt}
        }
}
\def\TAFs(#1,#2,#3,#4,#5,#6){
        \raisebox{-19.1pt}{
        \SetScale{0.5} \SetPFont{Helvetica}{14}
        \hspace{-15pt}
        \begin{picture}(50,39)(0,-4)
        \SetColor{Blue}
        \CArc(40,35)(25,90,270) \CArc(60,35)(25,270,90)
        \Line(40,60)(60,60) \Line(40,10)(60,10) \Line(50,10)(50,60)
        \Line(0,35)(15,35) \Line(85,35)(100,35)
        \SetColor{Black}
        \PText(53,38)(0)[l]{#5}
        \PText(40,65)(0)[rb]{#1} \PText(15,48)(0)[rb]{#6} \PText(65,62)(0)[lb]{#2}
        \PText(65,12)(0)[lt]{#3} \PText(35,12)(0)[rt]{#4}
        \SetColor{Red}
        \SetWidth{3}
                \Line(50,10)(50,60)
                \Line(40,60)(50,60)
                \CArc(40,35)(25,90,130)
                \Vertex(50,60){1.3}
        \SetWidth{0.5}
        \end{picture}
        \SetScale{1.0}
        \hspace{-7pt}
        }
}
\def\TAXs(#1,#2,#3,#4,#5,#6,#7){
        \raisebox{-19.1pt}{
        \SetScale{0.5} \SetPFont{Helvetica}{14}
        \hspace{-15pt}
        \begin{picture}(50,39)(0,-4)
        \SetColor{Blue}
        \CArc(40,35)(25,90,270) \CArc(60,35)(25,270,90)
        \Line(40,60)(60,60) \Line(40,10)(60,10) \Line(50,10)(50,60)
        \Line(0,35)(15,35) \Line(85,35)(100,35)
        \SetColor{Black}
        \PText(53,38)(0)[l]{#5}
        \PText(40,62)(0)[rb]{#1} \PText(15,48)(0)[rb]{#6} \PText(65,62)(0)[lb]{#2}
        \PText(65,12)(0)[lt]{#3} \PText(40,8)(0)[rt]{#4} \PText(15,22)(0)[rt]{#7}
        \SetColor{Red}
        \SetWidth{3}
                \CArc(40,35)(25,130,230)
        \SetWidth{0.5}
        \end{picture}
        \SetScale{1.0}
        \hspace{-7pt}
        }
}
\def\TAIs(#1,#2,#3,#4,#5,#6,#7){
        \raisebox{-19.1pt}{
        \SetScale{0.5} \SetPFont{Helvetica}{14}
        \hspace{-15pt}
        \begin{picture}(50,39)(0,-4)
        \SetColor{Blue}
        \CArc(40,35)(25,90,270) \CArc(60,35)(25,270,90)
        \Line(40,60)(60,60) \Line(40,10)(60,10) \Line(50,10)(50,60)
        \Line(0,35)(15,35) \Line(85,35)(100,35)
        \SetColor{Black}
        \PText(40,65)(0)[rb]{#1}
        \PText(60,65)(0)[lb]{#2}
        \PText(65,12)(0)[lt]{#3}
        \PText(35,12)(0)[rt]{#4}
        \PText(53,38)(0)[l]{#5}
        \PText(15,48)(0)[rb]{#6}
        \PText(85,48)(0)[lb]{#7}
        \SetColor{Red}
        \SetWidth{3}
                \Line(40,60)(60,60)
                \CArc(40,35)(25,90,130)
                \CArc(60,35)(25,50,90)
        \SetWidth{0.5}
        \end{picture}
        \SetScale{1.0}
        \hspace{-7pt}
        }
}

\title{FORM development}
\ShortTitle{FORM development}
\author{\speaker{J.A.M.Vermaseren}\\
 Nikhef, Science park 105, 1098XG, Amsterdam, The Netherlands \\
 E-mail: \email{t68@nikhef.nl}}
\abstract{I give an overview of FORM development based on a few pilot
projects, explaining how they have influenced the FORM capabilities. Next
I explain what is happnening right now in the field of Open Sourcing and
the FORM Forum.}
\FullConference{3rd Computational Particle Physics Workshop\\
  23rd- 25th September 2010 at KEK (Tsukuba)}

\begin{document}
 

\section{Introduction, Pilot Projects}

Development of FORM~\cite{Vermaseren:2000nd} has taken place mainly via 
what I call "pilot projects". These are science projects that are very 
demanding on algebraic systems and an efficient solution requires many new 
features. If FORM were to be a pure computer science undertaking, one would 
not have this and the result would be a product that is far less useful.
The main pilot projects are/have been:

\begin{itemize}
\item Three loop massless QCD (fixed moments).
\item The four loop beta function.
\item Three loop massless QCD in deep inelastic scattering (all moments).
\item The Karlsruhe projects.
\item Multiple Zeta Values.
\item Automatic One Loop calculations.
\end{itemize}

The first project started with the Mincer~\cite{Gorishnii:1989gt,
Larin:1991fz} program 
and the need for extreme speed. This led to special commands and functions. 
This project ran from 1990 till 1996.

The four loop beta function~\cite{beta3} led to the development of the 
color package~\cite{vanRitbergen:1998pn}. This required extensive treatment 
of antisymmetric functions and also some pattern recognition in the form of 
finding loops in index contractions. This project ran from 1996 till 1997.

The third project was more than just an extension of the first. It needed 
completely new techniques. This led to facilities for formal summation in 
the form of the Summer package~\cite{Vermaseren:1998uu} and large scale 
storage for tables. All needed several new features in FORM. The project 
ran basically from 1996 till 2005.

The need for computer power in Karlsruhe has led to the development of 
ParFORM~\cite{Fliegner:1999jq,Fliegner:2000uy,Tentyukov:2004hz}. This was 
mainly used for the 4-loop programs of Pavel Baikov~\cite{Baikov:2006ch}. 
But the concepts of the ParFORM program were largely taken over in TFORM 
and as such stand also at the cradle of that program. Another related 
development is the Laporta-style~\cite{LAPORTA} program that was developed 
in Karlsruhe and led to communication channels between FORM and other 
programs. The ParFORM project has been declared completed recently and ran 
from 1995 till 2010.

The Multiple Zeta Value~\cite{Blumlein:2009cf} calculations form a more 
mathematical project. They created the need to solve very large systems of 
equations and have been a major test case of TFORM~\cite{Tentyukov:2007mu}. 
It has led to completely new commands and new features in TFORM. This 
project ran from 1997 till 2010.

Since 2005 more attention is spent on the automated one loop calculations. 
This poses yet new requirements on FORM in the field of the manipulation of 
outputs and results. One can think here of factorization, code 
simplification and sophisticated print statements. This is the running 
project.

As part of this ongoing development it is of course important that 
developers have access to hardware that is really up-to-date and preferably 
more advanced than what the average user has at the same moment. This way 
the system will be ready for efficient use by the time that this hardware 
becomes more common. This is most noticeable with the parallel 
developments.

Example: TFORM~\cite{Tentyukov:2007mu} was developed on a machine with 4 
cores (Nikhef) in the days that everybody still had one core (or very 
rarely two). The past few years TFORM has been running mostly on eight 
cores (Karlsruhe and Zeuthen), and very recently Nikhef got a special 
computer for TFORM with 24 cores and 128 Gbytes of memory. By tuning TFORM 
more and more to such large numbers of cores, TFORM will be ready by the 
time everybody has access to such machines. At the moment a good TFORM 
computer, from the viewpoint of the user, would have 8 cores, a large 
memory (at least 32 Gbytes) and a very large and fast disk. And run 
LINUX~\footnote{On non-UNIX operating systems usually one or more features 
are missing. For instance the GMP does not work on Apple computers and 
Windows cannot handle the POSIX threads of TFORM.}.

In this talk we will shortly discuss a number of these pilot projects to 
see what they needed in (T)FORM and how they gave shape to it. The final 
project we discuss concerns the automated one loop calculations and their 
needs. This gives more insight in the future development of (T)FORM. 
Finally we will have a look at the most recent development: FORM is now 
open source and there is an internet forum for publicly discussing matters 
relating to FORM.


\section{Mincer}

When computing massless propagator graphs one can use integration by parts 
identities to reduce all one, two and three loop integrals to a set of 
three master integrals. These master integrals are known to sufficient 
powers in $\epsilon = (4-D)/2$ for the purpose of three loop (and even four 
loop) calculations.

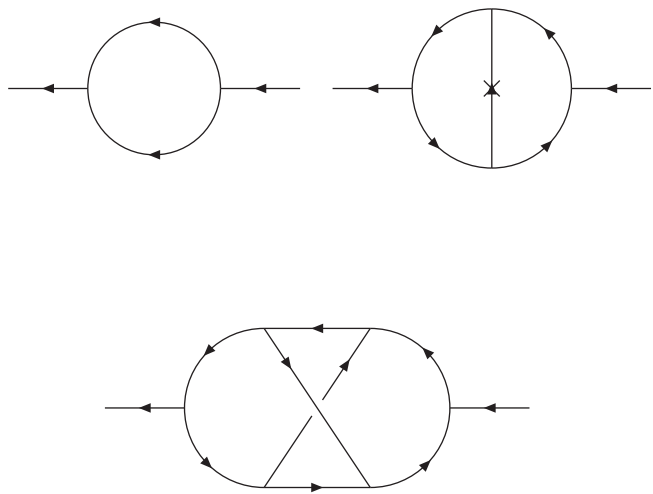
\begin{center}
\begin{picture}(130,110)(0,0)
\ArrowLine(40,60)(10,60)
\ArrowLine(120,60)(90,60)
\ArrowArc(65,60)(25,0,180)
\ArrowArcn(65,60)(25,0,180)
\end{picture}
\begin{picture}(120,110)(0,0)
\ArrowLine(30,60)(0,60)
\ArrowLine(120,60)(90,60)
\ArrowLine(60,30)(60,90)
\ArrowArc(60,60)(30,0,90)
\ArrowArc(60,60)(30,90,180)
\ArrowArc(60,60)(30,180,270)
\ArrowArc(60,60)(30,270,360)
\Line(57,57)(63,63)
\Line(57,63)(63,57)
\end{picture}
\begin{picture}(250,120)(0,0)
\ArrowLine(75,60)(45,60)
\ArrowLine(205,60)(175,60)
\ArrowLine(145,90)(105,90)
\ArrowLine(105,30)(145,30)
\ArrowLine(105,90)(123,63)
\ArrowLine(127,63)(145,90)
\Line(123,63)(145,30)
\Line(105,30)(123,57)
\ArrowArc(105,60)(30,90,180)
\ArrowArc(105,60)(30,180,270)
\ArrowArc(145,60)(30,270,360)
\ArrowArc(145,60)(30,0,90)
\end{picture}
\end{center}

To calculate higher Mellin moments of structure functions one has to 
consider scattering diagrams and take N derivatives with respect to the 
parton momentum P after which P is set to zero. These higher derivatives 
cause many tensorial and combinatorical problems and two functions 
(distrib\_ and dd\_) needed to be invented to deal with this properly. The 
strong point of especially dd\_ is that it gets the combinatorics right and 
terms are not generated multiple times.

Example:
{
\begin{verbatim}
    Vector Q,p1,p2,p3;
    Indices i1,...,i10;
    Tensor T;
    L	F1 = <Q(i1)>*...*<Q(i10)>;
    L	F2 = Q.p1^3*Q.p2^3*Q.p3^4;
    ToTensor,Q,T;
    Print;
    .sort

   F1 =
      T(i1,i2,i3,i4,i5,i6,i7,i8,i9,i10);

   F2 =
      T(p1,p1,p1,p2,p2,p2,p3,p3,p3,p3);

    id	T(?a) = dd_(?a);
    .sort

Time =       0.00 sec    Generated terms =        945
              F1         Terms in output =        945
                         Bytes used      =      33348

Time =       0.00 sec    Generated terms =          9
              F2         Terms in output =          9
                         Bytes used      =        572
    if ( expression(F1) )
    	Multiply p1(i1)*p1(i2)*p1(i3)
                *p2(i4)*p2(i5)*p2(i6)
    	        *p3(i7)*p3(i8)*p3(i9)*p3(i10);
    .end

Time =       0.00 sec    Generated terms =        945
              F1         Terms in output =          9
                         Bytes used      =        572

Time =       0.00 sec    Generated terms =          9
              F2         Terms in output =          9
                         Bytes used      =        572
\end{verbatim}
}

A function like dd\_ is also useful for one loop integration when one 
replaces
\begin{verbatim}
id  Q(i1?)*Q(i2?) = d_(i1,i2)*Q.Q/D
id  Q(i1?)*Q(i2?)*Q(i3?)*Q(i4?) =
           dd_(i1,i2,i3,i4)*Q.Q^2/D/(D+2)
id  Q(i1?)*Q(i2?)*Q(i3?)*Q(i4?)*Q(i5?)*Q(i6?) =
           dd_(i1,i2,i3,i4,i5,i6)*Q.Q^3/D/(D+2)/(D+4)
\end{verbatim}
Many of the other features that were introduced during the development and 
use of the Mincer~\cite{Gorishnii:1989gt,
Larin:1991fz} package are considered completely standard by now.


\section{Ensum}

The way N-dependent moments are computed is not by writing the derivatives 
out as sums and then working ones way through the Mincer algorithms, 
introducing more and more sums when the integration by parts identities are 
applied. This has been tried but only in the simplest two loop cases this 
has given results. In general this is too difficult.

The way that is used is by deriving recursion relations in the parameter 
N~\cite{Kazakov:1988jk} 
and then either summing the recursion, or when it is a higher order 
difference equation, solving it by brute force. This involves solving large 
sets of linear equations.

{
\begin{eqnarray}
  \TAF(1,1,1,1,E,n,1) & = & \frac{1}{\tilde{N}\plus 1\minus n}\ (
                \TAD(1,1,1,2,E,n)
                +\TAD(2,1,1,1,E,n)
                -\TAB(1,1,1,2,E,n) \nn
                +\frac{2P\mydot q}{q\mydot q}\ (
                \ (D\minus 4\minus\tilde{N}\minus E\plus n)\TAF(1,1,1,1,E,n,1)
					\nn
                +E\TAF(1,0,1,1,{E+1},n,1)
                -E\TAD(1,1,1,1,{E+1},n) \nn
                -\TAB(1,1,1,2,E,n)
                +n\TAD(1,1,1,1,E,{n-1})
                -n\TAB(1,1,1,1,E,{n-1}) 
        )\ ) \nonumber
\end{eqnarray}
}

An example of an integral to be solved by a difference equation is

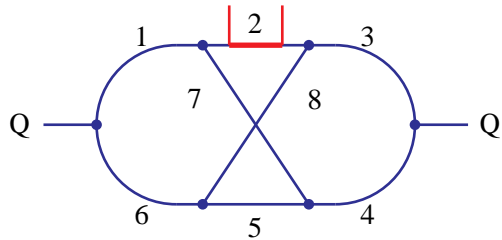
\begin{center}
\begin{picture}(200,100)(0,0)
	\SetWidth{1.0}
	\SetColor{Blue}
	\Line(70,80)(130,80)
	\Line(70,20)(130,20)
	\Line(80,20)(120,80)
	\Line(120,20)(80,80)
	\CArc(70,50)(30,90,270)
	\CArc(130,50)(30,270,90)
	\Line(20,50)(40,50)
	\Line(160,50)(180,50)
	\Vertex(40,50){2.0}
	\Vertex(160,50){2.0}
	\Vertex(80,20){2.0}
	\Vertex(120,20){2.0}
	\Vertex(80,80){2.0}
	\Vertex(120,80){2.0}
	\SetColor{Red}
	\SetWidth{1.0}
	\Line(90,80)(90,95)
	\Line(110,80)(110,95)
	\SetWidth{2.0}
	\Line(90,80)(110,80)
	\SetColor{Black}
	\SetWidth{1.0}
	\Text(15,50)[r]{Q}
	\Text(185,50)[l]{Q}
	\Text(60,80)[rb]{1}
	\Text(60,20)[rt]{6}
	\Text(140,80)[lb]{3}
	\Text(140,20)[lt]{4}
	\Text(100,85)[b]{2}
	\Text(100,15)[t]{5}
	\Text(80,60)[r]{7}
	\Text(120,60)[l]{8}
\end{picture}
\end{center}

\begin{eqnarray}
NO_{22}(N) \times ( \frac{2P\mydot Q}{Q\mydot Q} )^N & = &
\int d^Dp_1 d^D_2 d^Dp_3
\frac{(2P\mydot p_2)^N}{
	p_1^2\ (p_2^2)^{N+1}\ p_3^2\ p_4^2\ p_5^2\ p_6^2\ p_7^2\ p_8^2} \nonumber
\end{eqnarray}
 
Generically the equation looks like
\begin{eqnarray}
	a_0(N)\ F(N) + a_1(N)\ F(N-1) + \cdots + a_m(N)\ F(N-m)
			 & = & G(N) \nonumber
\end{eqnarray}
It is solved by making an ansatz containing many functions, substituting it 
and solving the resulting system of linear equations. One needs m fixed 
values for the boundary conditions.

For this diagram the equation is a third order equation with:
\begin{eqnarray}
	G(N) & = & 12 S_{-2}(N) ((-4N+2)\sign(N)+1) + 12 S_2(N)
				\nonumber \\ &&
		+24(1-\sign(N))/N-12(1-\sign(N))/N^2
				\nonumber \\
	a_0(N) & = & N(N-2\epsilon)(N+2\epsilon)
				(N+1-2\epsilon)(3N+3+2\epsilon)/2
				\nonumber \\
	a_1(N) & = & (N-2\epsilon)(15N\epsilon+4N\epsilon^3-3N-18N^2\epsilon
				\nonumber \\ &&
				-10N^2\epsilon^2+9N^2
				+5N^3\epsilon-9N^3+3N^4
				\nonumber \\ &&
				-2\epsilon+6\epsilon^2-8\epsilon^3+8\epsilon^4)/2
				\nonumber \\
	a_2(N) & = & (N-1)(12N\epsilon-28N\epsilon^2-160\epsilon^3
				\nonumber \\ &&
				-60N^2\epsilon+44N^2\epsilon^2
				+52N^3\epsilon+6N^3
				\nonumber \\ &&
				+6N^4+8\epsilon^2+56\epsilon^3-112\epsilon^4)/4
				\nonumber \\
	a_3(N) & = & (N-1)(N-2)(3N+2\epsilon)(N-1+3\epsilon)(N-1+6\epsilon)/2
				\nonumber
\end{eqnarray}

In the case of our diagram the answer is rather simple (this is 
exceptional):
\begin{eqnarray}
   F(N) & = & \theta(N)\frac{1+\sign(N)}{2}\frac{1}{1+N} (
			+20 \zeta_5
			+12 S_{-3,-2}(N+1)
				\nonumber \\ && \ \ \ \ \ \ \ \ \ \ 
			+4 S_{-3,2}(N+1)
			+8 S_{-2}(N+1) \zeta_3
			+4 S_{-2,-3}(N+1)
				\nonumber \\ && \ \ \ \ \ \ \ \ \ \ 
			-4 S_{-2,3}(N+1)
			+8 S_{2}(N+1) \zeta_3
			+4 S_{2,-3}(N+1)
				\nonumber \\ && \ \ \ \ \ \ \ \ \ \ 
			-4 S_{2,3}(N+1)
			+12 S_{3,-2}(N+1)
			+4 S_{3,2}(N+1)  ) \nonumber
\end{eqnarray}

Even then this turned out to be too demanding on the computers we had and 
it was needed to store all intermediately obtained integrals in a large set 
of tables.

The problem with tables is that they have to be compiled at the 
start of the program. Even at a few Mbytes/sec compiling 3 Gbytes of tables 
at the start of each program is not nice when you are developing new code. 

Hence a special database system for tables was designed: the 
tablebase~\cite{tablebase}. 
This has the tables in a special file (gzipped) and only tells FORM which 
elements there are. Then, when needed, only those elements that are 
actually used are compiled and applied. This turns out to work very well.

The whole made it possible to compute the anomalous dimensions and 
coefficient functions of three loop DIS in QCD~\cite{Moch:2004pa,
Vogt:2004mw,Vermaseren:2005qc}.


\section{Multiple Zeta Values}

Harmonic sums~\cite{Euler} are defined by~\cite{Vermaseren:1,HSUM3}:

\begin{eqnarray}
    S_m(N) & = & \sum_{i=1}^N \frac{1}{i^m} \nonumber \\
    S_{-m}(N) & = & \sum_{i=1}^N \frac{\sign(i)}{i^m} \nonumber \\
    S_{m,m_2,\cdots,m_p}(N) & = &
                \sum_{i=1}^N \frac{1}{i^m}S_{m_2,\cdots,m_p}(i) \nonumber 
\\
    S_{-m,m_2,\cdots,m_p}(N) & = &
                \sum_{i=1}^N \frac{\sign(m)}{i^m}S_{m_2,\cdots,m_p}(i)
                 \nonumber
\end{eqnarray}
This is a notation that is also suitable for computers. There is a 
difference here between various definitions as there are also people using 
$i-1$ for the argument of the $S$ in the recursive formula. Those sums we 
call $Z$-sums.

The harmonic polylogarithms~\cite{harmpol} are defined by:
\begin{eqnarray} 
  H(0;x) &=& \ln{x}        \nonumber \\
  H(1;x) &=& \int_0^x \frac{dx'}{1-x'} = - \ln(1-x) \nonumber \\
  H(-1;x) &=& \int_0^x \frac{dx'}{1+x'} = \ln(1+x)  \nonumber
\end{eqnarray} 
and the functions
$ f(0;x) = \frac{1}{x},\ \ \ \ 
  f(1;x) = \frac{1}{1-x},\ \ \ \ 
  f(-1;x) = \frac{1}{1+x}$

If $\vec{a}_w$ is an array with $w$ elements, all with value $a$, then:
\begin{eqnarray}
	H(\vec{0}_w;x) & = & \frac{1}{w!} \ln^w{x} \nonumber \\
	H(a,\vec{m}_w;x) & = & \int_0^x dx' \ f(a;x') \ H(\vec{m}_w;x') \nonumber
\end{eqnarray}

The weight is the number of indices in integral notation. These 
indices are either one or zero or minus one. The depth is the number of 
indices in sum notation in which there can be all integer numbers with the 
exception of zero. The sum of the absolute values of the indices in sum 
notation is equal to the weight. Harmonic sums are the Mellin transforms of 
the harmonic polylogarithms.

In the ensum project we needed these objects only to weight 6 and weight 5 
respectively. What was important was that we needed the harmonic 
polylogarithms in one (or the sums in infinity). There are many relations 
between them and because of that there are only very few that are linearly 
independent. This is very relevant as seen in the next example:

\begin{verbatim}
    #define SIZE "6"
    #include- harmpol.h
    Off statistics;
    .global
    Local F = S(R(-1,3,-2),N);
    #call invmel(S,N,H,x)
    Print +f +s;
    .end
    
   F =
       + H(R(-1,-3,0),x)*[1-x]^-1
       - 1/2*sign_(N)*H(R(1,0,0),x)*[1+x]^-1*z2
       + 1/2*H(R(-1,0,0),x)*[1-x]^-1*z2
       + 3/2*H(R(-1,0),x)*[1-x]^-1*z3
       + 21/20*H(R(-1),x)*[1-x]^-1*z2^2
       - 51/32*[1-x]^-1*z5
       + 3/4*[1-x]^-1*z2*z3
       - 7/2*s6
       + 51/32*z5*ln2
       - 33/64*z3^2
       + 9/4*z2*z3*ln2
       + 121/840*z2^3
       - 51/32*sign_(N)*[1+x]^-1*z5
       + 3/4*sign_(N)*[1+x]^-1*z2*z3
      ;
       
  0.28 sec out of 0.33 sec
\end{verbatim}
       
The above is a relatively 
short answer (14 terms). But this takes into account that there are many
relations between the harmonic sums in infinity (or the hpl's in one). If
we don't use these relations we have the result
\begin{verbatim}
       + H(R(-1,-3,0),x)*[1-x]^-1
       - sign_(N)*H(R(1,0,0),x)*Z(-2)*[1+x]^-1
       + H(R(-1,0,0),x)*Z(-2)*[1-x]^-1
       + 2*H(R(-1,0),x)*Z(-3)*[1-x]^-1
       + 3*H(R(-1),x)*Z(-4)*[1-x]^-1
       - sign_(N)*Z(-2,-3)*[1+x]^-1
       + 6*Z(-4,-1,1)      + 3*Z(-4,1,-1)
       + 5*Z(-3,-2,1)      + 4*Z(-3,-1,2)
       + Z(-3,1,-2)        + 3*Z(-3,2,-1)
       - Z(-2,-3)*[1-x]^-1 + 2*Z(-2,-3,-1)
       + 5*Z(-2,-3,1)      + Z(-2,-2,-2)
       + 3*Z(-2,-2,2)      + 2*Z(-2,-1,-3)
       + 2*Z(-2,-1,3)      + Z(-2,2,-2)
       + 3*Z(-2,3,-1)      + 3*Z(-1,-4,1)
       + 2*Z(-1,-3,2)      + Z(-1,-2,-3)
       + Z(-1,-2,3)        + Z(-1,3,-2)
       + 3*Z(-1,4,-1)
\end{verbatim}
Now we have 27 terms!
       
It is an interesting mathematical problem to see how many of these hpl's in 
one exist for a given weight. The only two ways know thus far to compute 
this are
\begin{itemize}
\item Determine a given object numerically to a very large number of 
digits. Guess a basis and evaluate the elements of this basis to the same 
accuracy. Then use a program like PSLQ or the LLL algorithm to determine an 
integer relation between them. This may or may not succeed, depending on 
the accuracy used.
\item Determine for a given weight all relations between the objects and 
solve this set. This can be done either as a matrix problem or formally 
with a computer algebra system. The power of the system determines how far 
one can go.
\end{itemize}

Although there exist formula's~\cite{Broadhurst:1,BK1} for the number of 
basis elements for given weight and depth, they have not been proven and 
sometimes surprises still show up (as happened in this research). The case 
of weight 27 was very special (a new phenomenon was expected to occur 
there) and finally solved (modulus a 31-bits prime number) recently in a 
job of 85 days on an 8-core Xeon computer at DESY Zeuthen~\cite{KV1}:
{\small
\begin{verbatim}
 171258.46 sec + 55845418.93 sec: 56016677.39 sec out of 7345664.84 sec
\end{verbatim}
}

As one can imagine, such calculations require optimal use of the hardware 
and several new features had to be added to (T)FORM. The effective use of 
the cores left only less than 5\% idle time during the whole job. This 
included occasional traffic jams at the single disk being used in 8 
parallel disk sorts.

Some of the new features~\cite{Vermaseren:2010iw} are
\begin{itemize}
\item The family of transform statements.
\item The InParallel option for TFORM to process large numbers of small 
expressions in parallel.
\item The use of the bracket index to divide the tasks over the workers.
\end{itemize}
And then there was the debugging of lots of features that had been used 
only rarely and hence were far from perfect.


\section{Automated One-Loop Calculations}

Originally FORM development started just for this problem. The name of the 
complete project was ESP (Experiment Simulation Program) and at the core of 
it a powerful symbolic manipulator was needed. The idea was to use an 
amplitude approach based on an advanced (at that moment) spinor library 
named Spider~\footnote{Of the spider approach only internal notes exist.} 
which had excellent numerical properties.

Hence in 1984 FORM development was started, but it took, of course, much 
longer than estimated and by the time it became operational (1989) the 
Grace~\cite{GRACE} system was well under development. Also I got 
sidetracked into three loop QCD to show off the power of FORM.

As a result the ESP system was never completed and it was judged wiser to 
join the Grace effort to reach the goal of automated one loop calculations.

But the project also resulted in the FF program by van 
Oldenborgh~\cite{vanOldenborgh:1,vanOldenborgh:2}.

One of the main problems in automated one loop calculations is 
organization. If the power of (T)FORM would not be sufficient, no other 
program would be able to deal with it. The main problem is the presentation 
of the output. The method used in the Grace system produces lengthy FORTRAN 
outputs and this in turns presents the FORTRAN compiler with unsurmountable 
complications. Hence the natural approach seems to be to try to make the 
outputs shorter by what is called code simplification. An example would be 
that
\begin{verbatim}
    F = x1*x3+x1*x4+x2*x3+x2*x4+x5
\end{verbatim}
is replaced by
\begin{verbatim}
    z1 = x1+x2
    z2 = x3+x4
    F = z1*z2+x5
\end{verbatim}
in which we save three multiplications and one addition.

Let us go to the current test reaction $e^-e^+\rightarrow\gamma e^-e^+$. 
There are two ways to attack this problem. The first way is to calculate 
the matrix element squared. This has been implemented~\cite{GraceForm} and 
a certain amount of simplification has been built in at the level of FORM 
code. This is rather slow and far from perfect. It gives an improvement of 
a factor between three and five. The whole reaction produces ${\cal 
O}(10^5)$ subroutines which, after improvement use $63\ 10^6$ additions and 
$70\ 10^6$
multiplications. The code can be compiled and made into a single 
executable, provided we use double precision. In quadruple precision the 
executable is too large (larger than 2 Gbytes) and the relocation mechanism 
of the GNU system is not up to the task.

Another way would be to compute the amplitude. This has advantages and 
disadvantages. The obvious disadvantage is that we have to deal with 
spinors and spin orientations. The advantage is a better numerical 
behaviour and an expression that is in principle linear in the number of 
diagrams. A sample input diagram is

\begin{center}
\begin{picture}(150,100)(30,30)
 \SetScale{1.5}
 \SetColor{Black}
  \ArrowLine(20,80)(50,80)   \Text(15,120)[]{1}
  \ArrowLine(50,20)(20,20)   \Text(15,30)[]{2}
  \ArrowLine(50,80)(50,50)
  \ArrowLine(50,50)(50,20)
  \ArrowLine(130,80)(100,80) \Text(210,120)[]{5}
  \ArrowLine(100,20)(130,20) \Text(210,30)[]{4}
  \ArrowLine(100,80)(100,20)
  \Vertex(50,80){1.5}
  \Vertex(100,80){1.5}
  \Vertex(50,20){1.5}
  \Vertex(100,20){1.5}
  \Vertex(50,50){1.5}
 \SetColor{Blue}
  \Photon(50,80)(100,80){3}{6} \Text(112.5,135)[]{$\gamma$}
  \Photon(100,20)(50,20){3}{6} \Text(112.5,15)[]{Z}
  \Photon(50,50)(80,50){3}{4}  \Text(130,75)[]{3}
\end{picture}
\end{center}

\begin{verbatim}

        -1
    *vfb(fl0,p2,amel)
    *ffvn(`czel1',`czel2',fl0,p2,l8,-l10,m8c)
    *sfn(fl0,l8,`amel')
    *ffvn(`cael1',`cael2',fl0,-l8,l6,-p3,n2a)
    *sfn(fl0,l6,`amel')
    *ffvn(`cael1',`cael2',fl0,-l6,p1,k7,m5c)
    *uf(fl0,p1,amel)
    *ufb(fl1,p4,amel)
    *ffvn(`czel1',`czel2',fl1,-p4,-l9,l10,m9c)
    *sfn(fl1,-l9,`amel')
    *ffvn(`cael1',`cael2',fl1,l9,-p5,-k7,m7c)
    *vf(fl1,p5,amel)
    *epsv(n2a,p3,ama)
    *dvn(m7c,m5c,k7,`ama')
    *dvn(m8c,m9c,l10,`amz')
    *num(2500)*loop(5)
    *mom1(q6,+p1)
    *mom1(q8,+p1-p3)
    *mom1(q9,+p5)
    *mom1(q10,+p4+p5)
    *mom2(2,l9,+q9+k7)
    *mom2(3,l10,+q10+k7)
    *mom2(4,l8,+q8+k7)
    *mom2(5,l6,+q6+k7)
    *mom3(k7,Q)
\end{verbatim}

We can see here the spinors. One way to deal with them is the `spider way', 
i.e. project them out onto the S,P,V,A,T currents and use the 10 spider 
relations to eliminate the tensor currents and contractions of the V and A 
currents with Levi-Civita tensors. When we bracket out the spinor and 
polarization vector dependent pieces there are `only' 580 different spin 
dependent objects that have to be computed 16 times. This means that we 
have to compute ${\cal O}(10^4)$ spin related quantities, compute 580 
scalar expressions and multiply those in. This is all very little compared 
to the millions of terms inside those 580 expressions.

Inside these expressions we have the loop integrals. We deal with them the 
`Grace way'~\cite{GRACE}. We can arrange in such a way that we have to 
compute each only once. There are in total 429 different loop integrals 
with their tensor structures. This in a total of 3456 diagrams of which 
3236 have a loop to be computed (the rest have counterterms). In contrast 
the matrix element squared method needs to calculate a loop integral 3236 
times.

At the moment we have no system of optimization yet and there are ${\cal 
O}(38\ 10^6)$ additions.and ${\cal O}(280\ 10^6)$ multiplications. The fact 
that already there are fewer additions gives good hope that after 
optimization this will be much shorter than the matrix element squared 
method as typical is about one multiplication per term after optimization.

Example:
{\footnotesize
\begin{verbatim}
 +L97(0)*(
  +9/16*amel^2*zk^2*inf*Z_79*Z_78*Z_76*Z_75*Z_74*Z_73^2*Z_69*Z_21*Z_1
  -15/8*amel^2*zk^2*inf*Z_79*Z_78*Z_76*Z_75*Z_74*Z_73^2*Z_69*Z_21*Z_12
  +9/4*amel^2*zk^2*inf*Z_79*Z_78*Z_76*Z_75*Z_74*Z_73^2*Z_69*Z_21*Z_13
  -amel^2*zk^2*inf*Z_79*Z_78*Z_76*Z_75*Z_74*Z_73^2*Z_69*Z_21*Z_15
  +9/16*amel^2*zk^2*inf*Z_79*Z_78*Z_77*Z_76*Z_75*Z_74*Z_73*Z_69*Z_21*Z_1
  -15/8*amel^2*zk^2*inf*Z_79*Z_78*Z_77*Z_76*Z_75*Z_74*Z_73*Z_69*Z_21*Z_12
  +9/4*amel^2*zk^2*inf*Z_79*Z_78*Z_77*Z_76*Z_75*Z_74*Z_73*Z_69*Z_21*Z_13
  -amel^2*zk^2*inf*Z_79*Z_78*Z_77*Z_76*Z_75*Z_74*Z_73*Z_69*Z_21*Z_15
  +9/16*amel^2*zk^2*inf*Z_80*Z_78*Z_77*Z_76*Z_75*Z_74*Z_73*Z_69*Z_21*Z_1
  -15/8*amel^2*zk^2*inf*Z_80*Z_78*Z_77*Z_76*Z_75*Z_74*Z_73*Z_69*Z_21*Z_12
  +9/4*amel^2*zk^2*inf*Z_80*Z_78*Z_77*Z_76*Z_75*Z_74*Z_73*Z_69*Z_21*Z_13
  -amel^2*zk^2*inf*Z_80*Z_78*Z_77*Z_76*Z_75*Z_74*Z_73*Z_69*Z_21*Z_15
  +9/16*amel^2*zk^2*inf*Z_80*Z_78*Z_77^2*Z_76*Z_75*Z_74*Z_69*Z_21*Z_1
  -15/8*amel^2*zk^2*inf*Z_80*Z_78*Z_77^2*Z_76*Z_75*Z_74*Z_69*Z_21*Z_12
  +9/4*amel^2*zk^2*inf*Z_80*Z_78*Z_77^2*Z_76*Z_75*Z_74*Z_69*Z_21*Z_13
  -amel^2*zk^2*inf*Z_80*Z_78*Z_77^2*Z_76*Z_75*Z_74*Z_69*Z_21*Z_15
  -9/32*amel^2*Ndim*zk^2*inf*Z_79*Z_78*Z_76*Z_75*Z_74*Z_73^2*Z_69*Z_21*Z_1
  +15/16*amel^2*Ndim*zk^2*inf*Z_79*Z_78*Z_76*Z_75*Z_74*Z_73^2*Z_69*Z_21*Z_12
  -9/8*amel^2*Ndim*zk^2*inf*Z_79*Z_78*Z_76*Z_75*Z_74*Z_73^2*Z_69*Z_21*Z_13
  +1/2*amel^2*Ndim*zk^2*inf*Z_79*Z_78*Z_76*Z_75*Z_74*Z_73^2*Z_69*Z_21*Z_15
  -9/32*amel^2*Ndim*zk^2*inf*Z_79*Z_78*Z_77*Z_76*Z_75*Z_74*Z_73*Z_69*Z_21*Z_1
  +15/16*amel^2*Ndim*zk^2*inf*Z_79*Z_78*Z_77*Z_76*Z_75*Z_74*Z_73*Z_69*Z_21*Z_12
  -9/8*amel^2*Ndim*zk^2*inf*Z_79*Z_78*Z_77*Z_76*Z_75*Z_74*Z_73*Z_69*Z_21*Z_13
  +1/2*amel^2*Ndim*zk^2*inf*Z_79*Z_78*Z_77*Z_76*Z_75*Z_74*Z_73*Z_69*Z_21*Z_15
  -9/32*amel^2*Ndim*zk^2*inf*Z_80*Z_78*Z_77*Z_76*Z_75*Z_74*Z_73*Z_69*Z_21*Z_1
  +15/16*amel^2*Ndim*zk^2*inf*Z_80*Z_78*Z_77*Z_76*Z_75*Z_74*Z_73*Z_69*Z_21*Z_12
  -9/8*amel^2*Ndim*zk^2*inf*Z_80*Z_78*Z_77*Z_76*Z_75*Z_74*Z_73*Z_69*Z_21*Z_13
  +1/2*amel^2*Ndim*zk^2*inf*Z_80*Z_78*Z_77*Z_76*Z_75*Z_74*Z_73*Z_69*Z_21*Z_15
  -9/32*amel^2*Ndim*zk^2*inf*Z_80*Z_78*Z_77^2*Z_76*Z_75*Z_74*Z_69*Z_21*Z_1
  +15/16*amel^2*Ndim*zk^2*inf*Z_80*Z_78*Z_77^2*Z_76*Z_75*Z_74*Z_69*Z_21*Z_12
  -9/8*amel^2*Ndim*zk^2*inf*Z_80*Z_78*Z_77^2*Z_76*Z_75*Z_74*Z_69*Z_21*Z_13
  +1/2*amel^2*Ndim*zk^2*inf*Z_80*Z_78*Z_77^2*Z_76*Z_75*Z_74*Z_69*Z_21*Z_15
    )
\end{verbatim}
}

This code has 32 terms and 412 multiplications but it is relatively easy to 
squeeze it to
\begin{verbatim}
  +L97(0)*zk^2*inf*amel^2*(Ndim-2)*(Z_79+Z_80)
   *Z_78*Z_77*(Z_77+Z_73)*Z_76*Z_75*Z_74*Z_69*Z_21
   *(-9*Z_1+30*Z_12-36*Z_13+16*Z_15)/32
\end{verbatim}
which involves 6 additions and 19 multiplications unless there are 
subexpressions that are common with other code in which case it is even 
less.

The object \verb:L97(0): is a scalar three-point function. The argument 
indicates which tensor integral is needed. We manage to store the powers of 
the various Feynman parameters in a single dimensional array in an optimal 
packing. This facilitates computing first all loop integrals and their 
tensor varieties and then using them from these arrays. This saves much 
time and space.

\subsection{Intermezzo}

If we have an N-point function, there can be at most N powers of the loop 
momentum in the numerator. This means that each Feynman parameter can have 
up to N powers and there are $N-1$ Feynman parameters. In an $N-1$ 
dimensional array there would be $(N+1)^{N-1}$ elements but actually we 
need only $\frac{(2N-1)!}{N!(N-1)!}$ elements. A good mapping for 
$x_1^{i_1}\cdots x_{N-1}^{i_{N-1}}$ to a single number K is
\begin{eqnarray}
	K_{i_1,\cdots,i_{N-1}} & = &
		B(2N-1,N)-1+\sum_{j=1}^{N-1} (-1)^j B(I_{N-j}-N,j) \nonumber \\
	I_j & = & \sum_{k=1}^j i_k \nonumber \\
	B(n,m) & = & \frac{n}{m} B(n-1,m-1)\ \ \ \ \ m > 0 \nonumber \\
	B(n,0) & = & 1 \nonumber
\end{eqnarray}
This can be programmed both in the FORM program and the FORTRAN program. In 
FORM it is much more compact.

It should be clear now that the code optimization is dominantly important. 
In the above example a simple factorization would suffice, but 
unfortunately that is usually not the case. We need techniques as used in 
compilers, but we have extra liberties. In a compiler one is not allowed to 
assume the addition to be associative or commutative. Here we can.

Of course the above compares are not completely fair. We have put the 
amplitude as a single expression in FORM, while the matrix element squared 
method worked diagram by diagram. We are however not so far that we can try 
to put that into FORM as a single expression. In the end the expression 
might be comparable in size, but especially in the early stages it would be 
much larger. There are other complications concerning D-dimensional indices 
versus 4-dimensional indices, because now the 4-dimensional indices can 
arrange themselves into loop-like structures and one has to keep them 
unsummed in the beginning at great cost. This is all much easier with the 
amplitudes as the only indices that can occur as in $\delta_{\alpha\alpha}$ 
are the loop indices. Everything outside the loop can be taken 
4-dimensional immediately.

At the moment work on code improvement and factorization is in an advanced 
stage, but not yet near completion. It will be interesting to see how much 
the expressions can be squeezed.


\section{Open Source}

Starting 26 Aug 2010 FORM has become open source. This means that there is 
a web based CVS from which anybody can download the sources of FORM and 
TFORM. There are some tools for configuration but because we have access 
only to a limited number of computers this is far from complete. Our hope 
is that users can make contributions here.

The license is the GNU Public License with the added hope that people will 
refer to the FORM publication when they use FORM for scientific 
publications.

The reason behind this move is that in a number of years, we do not know 
how many, FORM will have to survive without its original author. For this 
it is important that more people familiarize themselves with the sources 
and make additions. This can eventually only be done when the sources are 
generally available. Even so, it is not that easy to make additions to FORM 
because the code is more than 3.2 Mbytes (currently) (118000 lines) and not 
all of it is extensively documented. But there does exist much 
documentation if one compares it with similar programs. There is a 
testsuite based on the Ruby system, a layout program based on doxygen and 
of course there are lots of LaTeX files with explanations. For some program 
segments there is much commentary and for some (mostly older) segments 
there is unfortunately not very much commentary. Occasionally commentary is 
added, especially after a difficult debugging session.

Most of the work related to making FORM open source has been done by Jens 
Vollinga. This is fully in line with having more and more people involved 
with the development. The current drawback is that he will be leaving the 
academic environment. This may mean that he cannot spend much more time on 
FORM development (and GiNaC development).

Currently several people are working on new pieces of FORM code.

\begin{itemize}
\item Misha Tentyukov makes occasional additions as needed in Karlsruhe.
\item Jens Vollinga has made additions like systems independent .sav files.
\item Irina Pushkina works on code improvement for FORTRAN and/or C code.
\item Jan Kuipers works on rational polynomials, including factorization. 
If time is left in his contract he may create some facilities for Gr\"obner 
bases.
\item Thomas Reiter has put in most of the FORTRAN90 output mode.
\end{itemize}

In addition there are people who are very active in testing out new 
features and producing good bug reports. The importance of this should not 
be underestimated.


\section{The Forum}

To aid in dispersed development we (Jens Vollinga mainly) 
have set up a forum that allows people to communicate with each other. In 
principle this can be done without involvement of any of the main 
developers although, just in case, there will be moderators to remove 
inappropriate messages should they occur (like Spam).

The forum is located at $http://www.nikhef.nl/\sim form/forum$ and anybody 
can read it. To post messages you have to be a member. Subscription is 
rather easy.

For seeing how it works it is best to visit the site.


\section{Conclusions}

FORM development is slow work, but at the same time it makes steady 
progress.

Hopes are that the open source policy will add more impetus to this 
development.

Several projects are under way that will make outputs more compact.

 


\end{document}